\documentclass[pra,aps,preprint,showpacs]{revtex4}
\usepackage{epsfig,amssymb,amsfonts,amsmath}
\newcommand{\ket}[1]{|#1\rangle} \newcommand{\bra}[1]{\langle #1|}
\newcommand{\proj}[1]{\ket{#1}\bra{#1}}

\begin{document}

\title{Detecting a set of entanglement measures in an unknown tripartite quantum state by
local operations and classical communication}

\author{Yan-Kui Bai}
\affiliation{State Key Laboratory for Superlattices and
Microstructures, Institute of Semiconductors, Chinese Academy of
Sciences, P. O. Box 912, Beijing 100083, China}
\affiliation{Department of Physics \& Center of Theoretical and
Computational Physics, University of Hong Kong, Pokfulam Road,
Hong Kong, China}
\author{Shu-Shen Li and Hou-Zhi Zheng}
\affiliation{ CCAST (World Laboratory), P.O. Box 8730, Beijing
100080, China} \affiliation{State Key Laboratory for Superlattices
and Microstructures, Institute of Semiconductors, Chinese Academy
of Sciences, P. O. Box 912, Beijing 100083, China}
\author{Z. D. Wang}
\affiliation{Department of Physics \& Center of Theoretical and
Computational Physics, University of Hong Kong, Pokfulam Road, Hong
Kong, China}

\begin{abstract}
We propose a more general method for detecting a set of entanglement
measures, i.e. negativities, in an \emph{arbitrary} tripartite
quantum state by local operations and classical communication. To
accomplish the detection task using this method, three observers,
Alice, Bob and Charlie, do not need to perform the partial
transposition maps by the structural physical approximation;
instead, they are only required to collectively measure some
functions via three local networks supplemented by a classical
communication. With these functions, they are able to determine the
set of negativities related to the tripartite quantum state.
\end{abstract}

\pacs{03.67.Mn, 03.67.Lx, 03.67.Hk, 03.65.Ud}

\maketitle

\section{introduction}
Entanglement \cite{epr35} plays a vital role in quantum
information processing, such as quantum teleportation
\cite{ben93}, quantum key distribution \cite{eke91}, and quantum
dense code \cite{baw92}.

Before using the entanglement, one needs to make sure that it really
exists in a given system. For an unknown quantum state, one may
first perform the quantum state tomography
\cite{vor89,smi93,rwz05,blp05} which provides the full information
about the density matrix, and then evaluate the entanglement
property in terms of certain criterion and measure. However, the
quantum state tomography is not very efficient for the detection and
measurement of entanglement. Horodecki and Ekert proposed the direct
methods for detecting \cite{hae02} and measuring \cite{hol03}
entanglement in an unknown bipartite quantum state. Their idea is to
obtain the requisite eigenvalues by directly measuring some specific
functions of the unknown quantum state. For example, when checking
the positive partial transposition (PPT) criterion
\cite{per93,hhh96} in an two-qubit quantum state $\rho_{AB}$, the
observer can get the eigenvalues of the matrix $\rho_{AB}^{T_{A}}$
by directly measuring the functions
$\mbox{Tr}(\rho_{AB}^{T_{A}})^{k}$, for $k=2,3,4$ \cite{hae02}.
Comparing with the quantum state tomography, the direct method is
parametrically efficient. In Horodecki and Ekert's direct methods,
the structural physical approximation (SPA) technique \cite{pha03}
and a modified interferometer network \cite{spe00,ekl02} are
employed. Recently, Carteret proved that the SPA is unnecessary
\cite{car05,car06}, which makes the direct methods more feasible.
For multipartite entangled states, based on a set of entropic
inequalities, Alves \emph{et al.} put forward an efficient method
\cite{aaj04} for directly detecting entanglement in an optical
lattice. The implementation of the method is also analyzed
theoretically by Palmer \emph{et al} \cite{paj05}.

It is needful to characterize entanglement within local operations
and classical communication (LOCC) scenario. Curty \emph{et al.}
proved that entanglement is a precondition for secure quantum key
distribution \cite{cll04}. The LOCC schemes for directly detecting
and measuring entanglement in an unknown bipartite state have been
addressed in Refs. \cite{aho03,blj05,bla05}. In multiparty quantum
communication \cite{cgl99,cab02,czz05}, the multipartite entangled
state is an essential ingredient. Therefore the LOCC detection and
measurement of multipartite entanglement is worth to be considered.
The property of the multipartite entangled state can be
characterized partially by bipartite entanglement. For example, one
can detect the entanglement in a tripartite system in terms of a set
of PPT criteria, and furthermore, one can also quantify it with the
corresponding set of negativities \cite{vaw02}. Recently, Hyllus
\emph{et al.} designed an LOCC network for directly testing the PPT
criterion in a tripartite quantum state, which is assumed implicitly
to possess some specific symmetrical properties~\cite{hab04}.

In this paper, we generalize the network of Hyllus \emph{et al.} and
propose an LOCC method for detecting a set of negativities
\cite{vaw02} in the \emph{arbitrary} given tripartite quantum state.
Using this method, three observers, Alice, Bob, and Charlie, need
only to obtain the eigenvalues of a set of partial transposition
matrices via the generalized LOCC network, rather than to perform
the SPA. If the minimum eigenvalue of any partial transposition
matrix is negative, the tripartite quantum state is entangled and
the magnitude of entanglement can be measured in terms of the
corresponding negativity.

The paper is organized as follows. In Sec. II, we present in
detail the LOCC method for detecting negativities in an arbitrary
given tripartite quantum state. Then we discuss our method in Sec.
III. Finally, conclusions are given in Sec. IV.

\section{Detecting Negativities in an unknown tripartite quantum state by LOCC}
Negativity is a nontrivial entanglement measure, which is defined
as \cite{vaw02}
\begin{eqnarray}\label{1}
  \mathcal{N}(\rho) &=& \frac{\|\rho^{T_{A}}\|-1}{2},
\end{eqnarray}
where $\|\cdot\|$ denotes the trace norm which is the sum of the
moduli of eigenvalues for the hermitian matrix $\rho^{T_{A}}$.
This measure can be computed effectively for any mixed state of an
arbitrary bipartite system. Moreover, it gives an upper bound to
teleportation capacity.

The negativity can also be used to characterize the multipartite
entanglement. D\"{u}r \emph{et al.} suggested a useful way to
classify the entanglement properties of tripartite quantum state
$\rho_{ABC}$ by looking at the different bipartite splitting
\cite{dct99}. Therefore, one may use a set of negativities,
$\mathcal{N}_{(A-BC)}$, $\mathcal{N}_{(B-AC)}$,
$\mathcal{N}_{(C-AB)}$, $\mathcal{N}_{(A-B)}$, $\mathcal{N}_{(A-C)}$
and $\mathcal{N}_{(B-C)}$, to quantify the corresponding
entanglement in a tripartite system \cite{vaw02}.

We here develop an LOCC method to detecting the set of negativities
without performing the SPA. It is assumed that Alice, Bob and
Charlie share a number of copies of the unknown quantum state
$\rho_{ABC}$. The quantum state is defined on Hilbert space
$\mathcal{H=H_{A}\otimes H_{B}\otimes H_{C}}$ with the dimension as
$d=d_{A}\otimes d_{B}\otimes d_{C}$. The main task for the three
observers is to obtain the eigenvalues of the partial transposition
matrices $\rho^{T_{A}}_{ABC}$, $\rho^{T_{B}}_{ABC}$,
$\rho^{T_{C}}_{ABC}$, $\rho^{T_{A}}_{AB}$, $\rho^{T_{A}}_{AC}$ and
$\rho^{T_{B}}_{BC}$ within the LOCC scenario. A general LOCC network
used to accomplish this task is plotted in Fig.1, which is composed
of three local networks. The first part of Alice's local network is
a modified interferometer circuit (see \cite{hae02}; cf.
\cite{spe00,ekl02}) in which a controlled-$V_{k}$ gate is inserted.
Here, the function of the shift operator $V_{k}$ is \cite{ekl02}
\begin{eqnarray}\label{2}
 V_{k}\ket{\phi_{1}}\ket{\phi_{2}}\cdots\ket{\phi_{k}}
    =\ket{\phi_{k}}\ket{\phi_{1}}\cdots\ket{\phi_{k-1}}.
\end{eqnarray}
The second part is another interferometer circuit in which a
controlled-$R_{+}$ (or controlled-$R_{-}$) gate is inserted. The
hermitian and unitary operators $R_{+}$ and $R_{-}$ are defined as
\cite{bla05}
\begin{eqnarray}\label{3}
R_{+}=\frac{1}{\sqrt{2}}(\sigma_{z}+\sigma_{y})=\frac{1}{\sqrt{2}}\left(%
\begin{array}{cc}
  1 & -i \\
  i & -1 \\
\end{array}%
\right),
R_{-}=\frac{1}{\sqrt{2}}(\sigma_{z}-\sigma_{y})=\frac{1}{\sqrt{2}}\left(%
\begin{array}{cc}
  1 & i \\
  -i & -1 \\
\end{array}%
\right).
\end{eqnarray}
The local networks of Bob and Charlie are the same as that of Alice,
except for the different choices of the controlled operations in the
second part. (In fact, the first part of our LOCC network is just
the network proposed by Hyllus \emph{et al.}, see Fig.3 in Ref.
\cite{hab04}). In our LOCC method, Alice, Bob and Charlie can obtain
the eigenvalues of the set of partial transposition matrices by
making four groups of measurements. In the first group, they
implement the first part of the LOCC network and then measure the
output state of the ancillary qubits $a_{1}$, $b_{1}$ and $c_{1}$.
In other groups, they implement the whole LOCC network
and then measure the output state of the ancillary qubits $a_{2}$,
$b_{2}$ and $c_{2}$.

\begin{figure}
\begin{center} \epsfig{figure=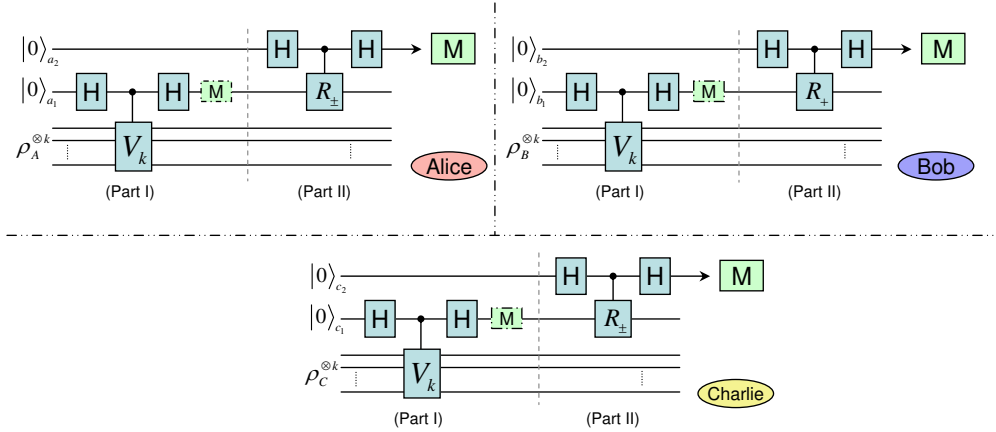,width=0.8\textwidth}
\caption{(Color online) A general network for remotely detecting
the negativities in an unknown tripartite quantum state.}
\end{center}
\end{figure}

Now we analyze the first part of the LOCC network. This part is
composed of three local modified interferometer circuits, in which
the Hadamard gate and the controlled-$V_{k}$ gate can be
represented by the unitary operators
\begin{eqnarray}\label{4}
H=\frac{1}{\sqrt{2}}\left(%
\begin{array}{cc}
  1 & 1 \\
  1 & -1 \\
\end{array}%
\right),
U_{C-V_{k}}=\left(%
\begin{array}{cc}
  1 & 0 \\
  0 & 0 \\
\end{array}%
\right)\otimes
I+\left(%
\begin{array}{cc}
  0 & 0 \\
  0 & 1 \\
\end{array}%
\right)\otimes V_{k},
\end{eqnarray}
respectively. In this part, the input state is
\begin{eqnarray}\label{5}
\rho_{in}(k)=\rho_{ABC}^{\otimes k}\otimes\rho_{a_{1}b_{1}c_{1}},
\end{eqnarray}
where the quantum state $\rho_{a_{1}b_{1}c_{1}}=\proj{000}$ is the
initial state of the ancillary qubits $a_{1}$, $b_{1}$ and
$c_{1}$. After passing through the three interferometer circuits,
the input state $\rho_{in}(k)$ will evolve into
\begin{eqnarray}\label{6}
  \rho_{out}'(k) &=&
  U_{h_{1}}U_{c-v}U_{h_{1}}\rho_{in}(k)U_{h_{1}}^{\dagger} U_{c-v}^{\dagger} U_{h_{1}}^{\dagger},
\end{eqnarray}
where $U_{h_{1}}=H_{a_{1}}\otimes H_{b_{1}}\otimes H_{c_{1}}\otimes
I_{ABC}^{\otimes k}$ and $U_{c-v}=U_{C_{a_{1}}-V_{Ak}}\otimes
U_{C_{b_{1}}-V_{Bk}}\otimes U_{C_{c_{1}}-V_{Ck}}$. In the output
state, what we care about is the quantum state evolution of
ancillary qubits $a_{1}$, $b_{1}$ and $c_{1}$. After tedious
derivations, the output state of the ancillary qubits is found to be
\begin{eqnarray}\label{7}
\rho_{a_{1}b_{1}c_{1}}'(k)&=&\mbox{Tr}_{ABC}[\rho_{out}'(k)]\nonumber\\
&=&\frac{1}{8}\left(%
\begin{array}{cccccccc}
  \mu_{1}^{(k)} & 0 & 0 & \mu_{9}^{(k)} & 0 & \mu_{10}^{(k)} & \mu_{11}^{(k)} & 0 \\
  0 & \mu_{2}^{(k)} & -\mu_{9}^{(k)} & 0 & -\mu_{10}^{(k)} & 0 & 0 & \mu_{12}^{(k)} \\
  0 & -\mu_{9}^{(k)} & \mu_{3}^{(k)} & 0 & -\mu_{11}^{(k)} & 0 & 0 & \mu_{13}^{(k)} \\
  \mu_{9}^{(k)} & 0 & 0 & \mu_{4}^{(k)} & 0 & -\mu_{12}^{(k)} & -\mu_{13}^{(k)} & 0 \\
  0 & -\mu_{10}^{(k)} & -\mu_{11}^{(k)} & 0 & \mu_{5}^{(k)} & 0 & 0 & \mu_{14}^{(k)} \\
  \mu_{10}^{(k)} & 0 & 0 & -\mu_{12}^{(k)} & 0 & \mu_{6}^{(k)} & -\mu_{14}^{(k)} & 0 \\
  \mu_{11}^{(k)} & 0 & 0 & -\mu_{13}^{(k)} & 0 & -\mu_{14}^{(k)} & \mu_{7}^{(k)} & 0 \\
  0 & \mu_{12}^{(k)} & \mu_{13}^{(k)} & 0 & \mu_{14}^{(k)} & 0 & 0 & \mu_{8}^{(k)} \\
\end{array}%
\right),
\end{eqnarray}
where
\begin{eqnarray*}
\mu_{1}^{(k)}&=&1+\alpha_{1}^{(k)}+\beta_{1}^{(k)}+\beta_{2}^{(k)}+\beta_{3}^{(k)}+\beta_{4}^{(k)}
+\beta_{5}^{(k)}+\beta_{6}^{(k)}+\gamma_{1}^{(k)}+\gamma_{2}^{(k)}+\gamma_{3}^{(k)}+\gamma_{4}^{(k)},\nonumber\\
\mu_{2}^{(k)}&=&1+\alpha_{2}^{(k)}+\beta_{1}^{(k)}-\beta_{2}^{(k)}-\beta_{3}^{(k)}+\beta_{4}^{(k)}
-\beta_{5}^{(k)}-\beta_{6}^{(k)}-\gamma_{1}^{(k)}-\gamma_{2}^{(k)}-\gamma_{3}^{(k)}-\gamma_{4}^{(k)},\nonumber\\
\mu_{3}^{(k)}&=&1+\alpha_{3}^{(k)}-\beta_{1}^{(k)}+\beta_{2}^{(k)}-\beta_{3}^{(k)}-\beta_{4}^{(k)}
+\beta_{5}^{(k)}-\beta_{6}^{(k)}-\gamma_{1}^{(k)}-\gamma_{2}^{(k)}-\gamma_{3}^{(k)}-\gamma_{4}^{(k)},\nonumber\\
\mu_{4}^{(k)}&=&1+\alpha_{4}^{(k)}-\beta_{1}^{(k)}-\beta_{2}^{(k)}+\beta_{3}^{(k)}-\beta_{4}^{(k)}
-\beta_{5}^{(k)}+\beta_{6}^{(k)}+\gamma_{1}^{(k)}+\gamma_{2}^{(k)}+\gamma_{3}^{(k)}+\gamma_{4}^{(k)},\nonumber
\end{eqnarray*}
\begin{eqnarray}\label{8}
\mu_{5}^{(k)}&=&1-\alpha_{4}^{(k)}-\beta_{1}^{(k)}-\beta_{2}^{(k)}+\beta_{3}^{(k)}-\beta_{4}^{(k)}
-\beta_{5}^{(k)}+\beta_{6}^{(k)}-\gamma_{1}^{(k)}-\gamma_{2}^{(k)}-\gamma_{3}^{(k)}-\gamma_{4}^{(k)},\nonumber\\
\mu_{6}^{(k)}&=&1-\alpha_{3}^{(k)}-\beta_{1}^{(k)}+\beta_{2}^{(k)}-\beta_{3}^{(k)}-\beta_{4}^{(k)}
+\beta_{5}^{(k)}-\beta_{6}^{(k)}+\gamma_{1}^{(k)}+\gamma_{2}^{(k)}+\gamma_{3}^{(k)}+\gamma_{4}^{(k)},\nonumber\\
\mu_{7}^{(k)}&=&1-\alpha_{2}^{(k)}+\beta_{1}^{(k)}-\beta_{2}^{(k)}-\beta_{3}^{(k)}+\beta_{4}^{(k)}
-\beta_{5}^{(k)}-\beta_{6}^{(k)}+\gamma_{1}^{(k)}+\gamma_{2}^{(k)}+\gamma_{3}^{(k)}+\gamma_{4}^{(k)},\nonumber\\
\mu_{8}^{(k)}&=&1-\alpha_{1}^{(k)}+\beta_{1}^{(k)}-\beta_{2}^{(k)}-\beta_{3}^{(k)}+\beta_{4}^{(k)}
-\beta_{5}^{(k)}-\beta_{6}^{(k)}-\gamma_{1}^{(k)}-\gamma_{2}^{(k)}-\gamma_{3}^{(k)}-\gamma_{4}^{(k)},\nonumber\\
\mu_{9}^{(k)}&=&\beta_{3}^{(k)}-\beta_{6}^{(k)}+\gamma_{1}^{(k)}+\gamma_{2}^{(k)}-\gamma_{3}^{(k)}-\gamma_{4}^{(k)},\nonumber\\
\mu_{10}^{(k)}&=&\beta_{2}^{(k)}-\beta_{5}^{(k)}+\gamma_{1}^{(k)}-\gamma_{2}^{(k)}+\gamma_{3}^{(k)}-\gamma_{4}^{(k)},\nonumber\\
\mu_{11}^{(k)}&=&\beta_{1}^{(k)}-\beta_{4}^{(k)}+\gamma_{1}^{(k)}-\gamma_{2}^{(k)}-\gamma_{3}^{(k)}+\gamma_{4}^{(k)},\nonumber\\
\mu_{12}^{(k)}&=&\beta_{1}^{(k)}-\beta_{4}^{(k)}-\gamma_{1}^{(k)}+\gamma_{2}^{(k)}+\gamma_{3}^{(k)}-\gamma_{4}^{(k)},\nonumber\\
\mu_{13}^{(k)}&=&\beta_{2}^{(k)}-\beta_{5}^{(k)}-\gamma_{1}^{(k)}+\gamma_{2}^{(k)}-\gamma_{3}^{(k)}+\gamma_{4}^{(k)},\nonumber\\
\mu_{14}^{(k)}&=&\beta_{3}^{(k)}-\beta_{6}^{(k)}-\gamma_{1}^{(k)}-\gamma_{2}^{(k)}+\gamma_{3}^{(k)}+\gamma_{4}^{(k)},
\end{eqnarray}

in which
\begin{eqnarray}\label{9}
\alpha_{1}^{(k)}&=&\mbox{Tr}[V_{Ak}\rho_{A}^{\otimes
k}]+\mbox{Tr}[V_{Bk}\rho_{B}^{\otimes
k}]+\mbox{Tr}[V_{Ck}\rho_{C}^{\otimes
k}]=\mbox{Tr}\rho_{A}^{k}+\mbox{Tr}\rho_{B}^{k}+\mbox{Tr}\rho_{C}^{k},\nonumber\\
\alpha_{2}^{(k)}&=&\mbox{Tr}[V_{Ak}\rho_{A}^{\otimes
k}]+\mbox{Tr}[V_{Bk}\rho_{B}^{\otimes
k}]-\mbox{Tr}[V_{Ck}\rho_{C}^{\otimes
k}]=\mbox{Tr}\rho_{A}^{k}+\mbox{Tr}\rho_{B}^{k}-\mbox{Tr}\rho_{C}^{k},\nonumber\\
\alpha_{3}^{(k)}&=&\mbox{Tr}[V_{Ak}\rho_{A}^{\otimes
k}]-\mbox{Tr}[V_{Bk}\rho_{B}^{\otimes
k}]+\mbox{Tr}[V_{Ck}\rho_{C}^{\otimes
k}]=\mbox{Tr}\rho_{A}^{k}-\mbox{Tr}\rho_{B}^{k}+\mbox{Tr}\rho_{C}^{k},\nonumber\\
\alpha_{4}^{(k)}&=&\mbox{Tr}[V_{Ak}\rho_{A}^{\otimes
k}]-\mbox{Tr}[V_{Bk}\rho_{B}^{\otimes
k}]-\mbox{Tr}[V_{Ck}\rho_{C}^{\otimes
k}]=\mbox{Tr}\rho_{A}^{k}-\mbox{Tr}\rho_{B}^{k}-\mbox{Tr}\rho_{C}^{k},\nonumber\\
\beta_{1}^{(k)}&=&\frac{1}{2}\mbox{Tr}[(V_{Ak}\otimes
V_{Bk})\rho_{AB}^{\otimes k}]=\frac{1}{2}\mbox{Tr}\rho_{AB}^{k},\nonumber\\
\beta_{2}^{(k)}&=&\frac{1}{2}\mbox{Tr}[(V_{Ak}\otimes
V_{Ck})\rho_{AC}^{\otimes k}]=\frac{1}{2}\mbox{Tr}\rho_{AC}^{k},\nonumber\\
\beta_{3}^{(k)}&=&\frac{1}{2}\mbox{Tr}[(V_{Bk}\otimes
V_{Ck})\rho_{BC}^{\otimes k}]=\frac{1}{2}\mbox{Tr}\rho_{BC}^{k},\nonumber\\
\beta_{4}^{(k)}&=&\frac{1}{2}\mbox{Tr}[(V_{Ak}^{\dagger}\otimes
V_{Bk})\rho_{AB}^{\otimes k}]=\frac{1}{2}\mbox{Tr}(\rho_{AB}^{T_{A}})^{k},\nonumber\\
\beta_{5}^{(k)}&=&\frac{1}{2}\mbox{Tr}[(V_{Ak}^{\dagger}\otimes
V_{Ck})\rho_{AC}^{\otimes k}]=\frac{1}{2}\mbox{Tr}(\rho_{AC}^{T_{A}})^{k},\nonumber\\
\beta_{6}^{(k)}&=&\frac{1}{2}\mbox{Tr}([V_{Bk}^{\dagger}\otimes
V_{Ck})\rho_{BC}^{\otimes
k}]=\frac{1}{2}\mbox{Tr}(\rho_{BC}^{T_{B}})^{k},\nonumber\\
\gamma_{1}^{(k)}&=&\frac{1}{4}\mbox{Tr}[(V_{Ak}\otimes
V_{Bk}\otimes V_{Ck})\rho_{ABC}^{\otimes k}]=\frac{1}{4}\mbox{Tr}\rho_{ABC}^{k},\nonumber\\
\gamma_{2}^{(k)}&=&\frac{1}{4}\mbox{Tr}[(V_{Ak}^{\dagger}\otimes
V_{Bk}\otimes V_{Ck})\rho_{ABC}^{\otimes
k}]=\frac{1}{4}\mbox{Tr}(\rho_{ABC}^{T_{A}})^{k},\nonumber\\
\gamma_{3}^{(k)}&=&\frac{1}{4}\mbox{Tr}[(V_{Ak}\otimes
V_{Bk}^{\dagger}\otimes V_{Ck})\rho_{ABC}^{\otimes
k}]=\frac{1}{4}\mbox{Tr}(\rho_{ABC}^{T_{B}})^{k},\nonumber\\
\gamma_{4}^{(k)}&=&\frac{1}{4}\mbox{Tr}[(V_{Ak}\otimes
V_{Bk}\otimes V_{Ck}^{\dagger})\rho_{ABC}^{\otimes
k}]=\frac{1}{4}\mbox{Tr}(\rho_{ABC}^{T_{C}})^{k}.
\end{eqnarray}
In the derivation of Eq. \eqref{7}, we made use of the cyclicity
of the trace and the property
$\mbox{Tr}U^{\dagger}\rho=(\mbox{Tr}U \rho)^{\ast}$. In Eq.
\eqref{9}, the relations between the parameters
$\alpha_{i}^{(k)}$, $\beta_{i}^{(k)}$, $\gamma_{i}^{(k)}$ and the
traces of the corresponding matrices were analyzed in Ref.
\cite{ekl02,aho03,car05,bla05,hab04}. In the first group of
measurements, Alice, Bob and Charlie measure the expectation
values of $\sigma_{z}\otimes\sigma_{z}\otimes\sigma_{z}$ on the
output state $\rho_{a_{1}b_{1}c_{1}}'(k)$, for $k=2,3,\cdots,d$.
With these expectation values, they can get
\begin{eqnarray}\label{10}
\mbox{Tr}[(\sigma_{z}\otimes\sigma_{z}\otimes\sigma_{z})\rho_{a_{1}b_{1}c_{1}}'(k)]
&=&\gamma_{1}^{(k)}+\gamma_{2}^{(k)}+\gamma_{3}^{(k)}+\gamma_{4}^{(k)}\nonumber\\
&=&\frac{1}{4}\mbox{Tr}\left(\rho_{ABC}^{k}+
(\rho_{ABC}^{T_{A}})^{k}+(\rho_{ABC}^{T_{B}})^{k}+(\rho_{ABC}^{T_{C}})^{k}\right).
\end{eqnarray}
These expectation values can be obtained by collectively measuring
the probabilities $P_{a_{1}b_{1}c_{1}}^{(k)}(ijl)$ of the output
state $\rho_{a_{1}b_{1}c_{1}}'(k)$ being found in the states
$\ket{000}$,$\ket{001}$,$\ket{010}$,$\ket{011}$,$\ket{100}$,$\ket{101}$,$\ket{110}$
and $\ket{111}$, respectively, in which a classical communication
is needed. With the probabilities
$P_{a_{1}b_{1}c_{1}}^{(k)}(ijl)$, the three observers can also get
the following relations:
\begin{eqnarray}
\label{11}
\mbox{Tr}[(\sigma_{z}\otimes\sigma_{z})\rho_{a_{1}b_{1}}'(k)]
&=&\beta_{1}^{(k)}+\beta_{4}^{(k)}=
\frac{1}{2}\left(\mbox{Tr}\rho_{AB}^{k}+\mbox{Tr}(\rho_{AB}^{T_{A}})^{k}\right)\nonumber\\
\mbox{Tr}[(\sigma_{z}\otimes\sigma_{z})\rho_{a_{1}c_{1}}'(k)]
&=&\beta_{2}^{(k)}+\beta_{5}^{(k)}=
\frac{1}{2}\left(\mbox{Tr}\rho_{AC}^{k}+\mbox{Tr}(\rho_{AC}^{T_{A}})^{k}\right)\nonumber\\
\mbox{Tr}[(\sigma_{z}\otimes\sigma_{z})\rho_{b_{1}c_{1}}'(k)]
&=&\beta_{3}^{(k)}+\beta_{6}^{(k)}=
\frac{1}{2}\left(\mbox{Tr}\rho_{BC}^{k}+\mbox{Tr}(\rho_{BC}^{T_{B}})^{k}\right),
\end{eqnarray}
where $\rho_{a_{1}b_{1}}'(k)$, $\rho_{a_{1}c_{1}}'(k)$ and
$\rho_{b_{1}c_{1}}'(k)$ are the bipartite reduced density matrices
of $\rho_{a_{1}b_{1}c_{1}}'(k)$. In Eqs. (10) and (11), the case
for $k=2$ is special, which is due to the hermitian property of
the shift operator $V_{2}$.  Combining the hermitian property of
$V_{2}$ and the definitions of $\beta_{i}$, $\gamma_{i}$, the
three observers can have the following relations
\begin{eqnarray}\label{12}
&&\mbox{Tr}[(\sigma_{z}\otimes\sigma_{z}\otimes\sigma_{z})\rho_{a_{1}b_{1}c_{1}}'(2)]=
\mbox{Tr}\rho_{ABC}^{2}=\mbox{Tr}(\rho_{ABC}^{T_{A}})^{2}=\mbox{Tr}(\rho_{ABC}^{T_{B}})^{2}
=\mbox{Tr}(\rho_{ABC}^{T_{C}})^{2}\nonumber\\
&&\mbox{Tr}[(\sigma_{z}\otimes\sigma_{z})\rho_{a_{1}b_{1}}'(2)]=\mbox{Tr}\rho_{AB}^{2}
=\mbox{Tr}(\rho_{AB}^{T_{A}})^{2}\nonumber\\
&&\mbox{Tr}[(\sigma_{z}\otimes\sigma_{z})\rho_{a_{1}c_{1}}'(2)]=\mbox{Tr}\rho_{AC}^{2}
=\mbox{Tr}(\rho_{AC}^{T_{A}})^{2}\nonumber\\
&&\mbox{Tr}[(\sigma_{z}\otimes\sigma_{z})\rho_{b_{1}c_{1}}'(2)]=\mbox{Tr}\rho_{BC}^{2}
=\mbox{Tr}(\rho_{BC}^{T_{B}})^{2}.
\end{eqnarray}
While, for $k>2$, the shift operator $V_{k}$ is not hermitian
\cite{pha03}, and the three observers cannot have the same relations
as Eq. \eqref{12}. From the above analysis, we can see that the
three observers cannot obtain the requisite eigenvalues in general,
unless the tripartite quantum state has the symmetrical property
$\mbox{Tr}\rho_{ABC}^{k}=\mbox{Tr}(\rho_{ABC}^{T_{A}})^{k}=\mbox{Tr}(\rho_{ABC}^{T_{B}})^{k}
=\mbox{Tr}(\rho_{ABC}^{T_{C}})^{k}$. This point seems to have been
neglected in Ref. \cite{hab04}.

In order to obtain the eigenvalues of the set of partial
transposition matrices for an \emph{arbitrary} given tripartite
quantum state, Alice, Bob and Charlie need to make other
measurements. In the second group of measurements, they need to
implement the whole LOCC network shown in Fig.1, in which they
choose the controlled gates in the second part to be
controlled-$R_{-}$, controlled-$R_{+}$ and controlled-$R_{+}$,
respectively.
In the second part, the input state is
\begin{eqnarray}\label{13}
  \rho_{in}'(k) &=&
  \rho'_{a_{1}b_{1}c_{1}}(k)\otimes\rho_{a_{2}b_{2}c_{2}},
\end{eqnarray}
where the quantum state $\rho_{a_{2}b_{2}c_{2}}=\proj{000}$ is the
initial state of the ancillary qubits $a_{2}$, $b_{2}$ and
$c_{2}$. Passing through the three interferometer circuits, the
input state $\rho_{in}'(k)$ will evolve into the following form:
\begin{eqnarray}\label{14}
  \rho_{out}^{-++}(k) &=&
  U_{h_{2}}U_{c-r_{1}}U_{h_{2}}\rho'_{in}(k)U_{h_{2}}^{\dagger}U_{c-r_{1}}^{\dagger}U_{h_{2}}^{\dagger},
\end{eqnarray}
where $U_{h_{2}}=H_{a_{2}}\otimes H_{b_{2}}\otimes
H_{c_{2}}\otimes I_{a_{1}b_{1}c_{1}}$ and
$U_{c-r_{1}}=U_{C_{a_{2}}-R_{-}}\otimes U_{C_{b_{2}}-R_{+}}\otimes
U_{C_{c_{2}}-R_{+}}$. In the output state $\rho_{out}^{-++}(k)$,
what we care about is the quantum state evolution of the ancillary
qubits $a_{2}$, $b_{2}$ and $c_{2}$. The corresponding output
state is found to be
\begin{eqnarray}\label{15}
  \rho_{a_{2}b_{2}c_{2}}^{-++}(k) &=&
  \mbox{Tr}_{a_{1}b_{1}c_{1}}[\rho_{out}^{-++}(k)]\nonumber\\
&=&\frac{1}{8}\left(%
\begin{array}{cccccccc}
  \nu_{1}^{(k)} & 0 & 0 & 0 & 0 & 0 & 0 & 0 \\
  0 & \nu_{2}^{(k)} & 0 & 0 & 0 & 0 & 0 & 0 \\
  0 & 0 & \nu_{3}^{(k)} & 0 & 0 & 0 & 0 & 0 \\
  0 & 0 & 0 & \nu_{4}^{(k)} & 0 & 0 & 0 & 0 \\
  0 & 0 & 0 & 0 & \nu_{5}^{(k)} & 0 & 0 & 0 \\
  0 & 0 & 0 & 0 & 0 & \nu_{6}^{(k)} & 0 & 0 \\
  0 & 0 & 0 & 0 & 0 & 0 & \nu_{7}^{(k)} & 0 \\
  0 & 0 & 0 & 0 & 0 & 0 & 0 & \nu_{8}^{(k)} \\
\end{array}%
\right),
\end{eqnarray}
where
\begin{eqnarray}\label{16}
\nu_{1}^{(k)}&=&1+\frac{1}{\sqrt{2}}\alpha_{1}^{(k)}+\beta_{1}^{(k)}+\beta_{2}^{(k)}+\beta_{6}^{(k)}
+\frac{1}{\sqrt{2}}\Gamma^{(k)},\nonumber\\
\nu_{2}^{(k)}&=&1+\frac{1}{\sqrt{2}}\alpha_{2}^{(k)}+\beta_{1}^{(k)}-\beta_{2}^{(k)}-\beta_{6}^{(k)}
-\frac{1}{\sqrt{2}}\Gamma^{(k)},\nonumber\\
\nu_{3}^{(k)}&=&1+\frac{1}{\sqrt{2}}\alpha_{3}^{(k)}-\beta_{1}^{(k)}+\beta_{2}^{(k)}-\beta_{6}^{(k)}
-\frac{1}{\sqrt{2}}\Gamma^{(k)},\nonumber\\
\nu_{4}^{(k)}&=&1+\frac{1}{\sqrt{2}}\alpha_{4}^{(k)}-\beta_{1}^{(k)}-\beta_{2}^{(k)}+\beta_{6}^{(k)}
+\frac{1}{\sqrt{2}}\Gamma^{(k)},\nonumber
\end{eqnarray}
\begin{eqnarray}
\nu_{5}^{(k)}&=&1-\frac{1}{\sqrt{2}}\alpha_{4}^{(k)}-\beta_{1}^{(k)}-\beta_{2}^{(k)}+\beta_{6}^{(k)}
-\frac{1}{\sqrt{2}}\Gamma^{(k)},\nonumber\\
\nu_{6}^{(k)}&=&1-\frac{1}{\sqrt{2}}\alpha_{3}^{(k)}-\beta_{1}^{(k)}+\beta_{2}^{(k)}-\beta_{6}^{(k)}
+\frac{1}{\sqrt{2}}\Gamma^{(k)},\nonumber\\
\nu_{7}^{(k)}&=&1-\frac{1}{\sqrt{2}}\alpha_{2}^{(k)}+\beta_{1}^{(k)}-\beta_{2}^{(k)}-\beta_{6}^{(k)}
+\frac{1}{\sqrt{2}}\Gamma^{(k)},\nonumber\\
\nu_{8}^{(k)}&=&1-\frac{1}{\sqrt{2}}\alpha_{1}^{(k)}+\beta_{1}^{(k)}+\beta_{2}^{(k)}+\beta_{6}^{(k)}
-\frac{1}{\sqrt{2}}\Gamma^{(k)},
\end{eqnarray}
in which
$\Gamma^{(k)}=\gamma_{1}^{(k)}-\gamma_{2}^{(k)}+\gamma_{3}^{(k)}+\gamma_{4}^{(k)}$.
In the second group of measurements, Alice, Bob and Charlie measure
the expectation values of
$\sigma_{z}\otimes\sigma_{z}\otimes\sigma_{z}$ on the output state
$\rho_{a_{2}b_{2}c_{2}}^{-++}(k)$ for $k=3,4,\cdots,d$. These
expectation values can be written as
\begin{eqnarray}\label{17}
\mbox{Tr}[(\sigma_{z}\otimes\sigma_{z}\otimes\sigma_{z})\rho_{a_{2}b_{2}c_{2}}^{-++}(k)]&=&\frac{1}{\sqrt{2}}
(\gamma_{1}^{(k)}-\gamma_{2}^{(k)}+\gamma_{3}^{(k)}+\gamma_{4}^{(k)})\nonumber\\
&=& \frac{1}{4\sqrt{2}}\mbox{Tr}\left[\rho_{ABC}^{k}-
(\rho_{ABC}^{T_{A}})^{k}+(\rho_{ABC}^{T_{B}})^{k}+(\rho_{ABC}^{T_{C}})^{k}\right].
\end{eqnarray}
When they consider the expectation values of the bipartite reduced
density matrices of $\rho_{a_{2}b_{2}c_{2}}^{-++}(k)$, they have
\begin{eqnarray}\label{18}
\mbox{Tr}[(\sigma_{z}\otimes\sigma_{z})\rho_{a_{2}b_{2}}^{-++}(k)]
&=&\beta_{1}^{(k)}=\frac{1}{2}\mbox{Tr}\rho_{AB}^{k}\nonumber\\
\mbox{Tr}[(\sigma_{z}\otimes\sigma_{z})\rho_{a_{2}c_{2}}^{-++}(k)]
&=&\beta_{2}^{(k)}=\frac{1}{2}\mbox{Tr}\rho_{AC}^{k}\nonumber\\
\mbox{Tr}[(\sigma_{z}\otimes\sigma_{z})\rho_{b_{2}c_{2}}^{-++}(k)]
&=&\beta_{6}^{(k)}=\frac{1}{2}\mbox{Tr}(\rho_{BC}^{T_{B}})^{k}.
\end{eqnarray}
The expectation values in Eq. \eqref{17} and Eq. \eqref{18} can be
obtain by measuring the probabilities
$P_{a_{2}b_{2}c_{2}}^{(-++)(k)}(ijl)$ of the output state
$\rho_{a_{2}b_{2}c_{2}}^{-++}(k)$ being found in the states
$\{\ket{ijl}\}$.

In the third group of measurements, Alice, Bob and Charlie
implement again the whole LOCC network. This time, they choose the
controlled gates in the second part to be controlled-$R_{-}$,
controlled-$R_{+}$ and controlled-$R_{-}$, respectively. The
output state of the ancillary qubits $a_{2}$, $b_{2}$ and $c_{2}$
is
\begin{eqnarray}\label{19}
\rho_{a_{2}b_{2}c_{2}}^{-+-}(k)=\mbox{Tr}_{a_{1}b_{1}c_{1}}[\rho_{out}^{-+-}(k)]
&=&\mbox{Tr}_{a_{1}b_{1}c_{1}}[U_{h_{2}}U_{c-r_{2}}U_{h_{2}}\rho'_{in}(k)U_{h_{2}}
^{\dagger}U_{c-r_{2}}^{\dagger}U_{h_{2}}^{\dagger}],
\end{eqnarray}
where $U_{c-r_{2}}=U_{C_{a_{2}}-R_{-}}\otimes
U_{C_{b_{2}}-R_{+}}\otimes U_{C_{c_{2}}-R_{-}}$. By measuring the
probabilities $P_{a_{2}b_{2}c_{2}}^{(-+-)(k)}(ijl)$ of the output
state $\rho_{a_{2}b_{2}c_{2}}^{-+-}(k)$ being found in the states
$\{\ket{ijl}\}$, they can obtain the following relations
\begin{eqnarray}\label{20}
\mbox{Tr}[(\sigma_{z}\otimes\sigma_{z}\otimes\sigma_{z})\rho_{a_{2}b_{2}c_{2}}^{-+-}(k)]&=&\frac{1}{\sqrt{2}}
(\gamma_{1}^{(k)}+\gamma_{2}^{(k)}-\gamma_{3}^{(k)}+\gamma_{4}^{(k)})\nonumber\\
&=&\frac{1}{4\sqrt{2}}\mbox{Tr}\left(\rho_{ABC}^{k}+
(\rho_{ABC}^{T_{A}})^{k}-(\rho_{ABC}^{T_{B}})^{k}+(\rho_{ABC}^{T_{C}})^{k}\right),\nonumber\\
\mbox{Tr}[(\sigma_{z}\otimes\sigma_{z})\rho_{a_{2}b_{2}}^{-+-}(k)]
&=&\beta_{1}^{(k)}=\frac{1}{2}\mbox{Tr}\rho_{AB}^{k}\nonumber\\
\mbox{Tr}[(\sigma_{z}\otimes\sigma_{z})\rho_{a_{2}c_{2}}^{-+-}(k)]
&=&\beta_{5}^{(k)}=\frac{1}{2}\mbox{Tr}(\rho_{AC}^{k})^{T_{A}}\nonumber\\
\mbox{Tr}[(\sigma_{z}\otimes\sigma_{z})\rho_{b_{2}c_{2}}^{-+-}(k)]
&=&\beta_{3}^{(k)}=\frac{1}{2}\mbox{Tr}\rho_{BC}^{k},
\end{eqnarray}
where $k=3,4,\cdots,d$.

In the fourth group of measurements, the three observers still
implement the whole LOCC network, but at this time they choose the
controlled gates in the second part to be controlled-$R_{+}$,
controlled-$R_{+}$ and controlled-$R_{-}$, respectively. The
output state of the ancillary qubits $a_{2}$, $b_{2}$ and $c_{2}$
is
\begin{eqnarray}\label{21}
\rho_{a_{2}b_{2}c_{2}}^{++-}(k)=\mbox{Tr}_{a_{1}b_{1}c_{1}}[\rho_{out}^{++-}(k)]
&=&\mbox{Tr}_{a_{1}b_{1}c_{1}}[U_{h_{2}}U_{c-r_{3}}U_{h_{2}}\rho'_{in}(k)U_{h_{2}}
^{\dagger}U_{c-r_{3}}^{\dagger}U_{h_{2}}^{\dagger}],
\end{eqnarray}
where $U_{c-r_{3}}=U_{C_{a_{2}}-R_{+}}\otimes
U_{C_{b_{2}}-R_{+}}\otimes U_{C_{c_{2}}-R_{-}}$. By measuring the
probabilities $P_{a_{2}b_{2}c_{2}}^{(++-)(k)}(ijl)$ of the output
state $\rho_{a_{2}b_{2}c_{2}}^{++-}(k)$ being found in the states
$\{\ket{ijl}\}$, Alice, Bob and Charlie have
\begin{eqnarray}\label{22}
\mbox{Tr}[(\sigma_{z}\otimes\sigma_{z}\otimes\sigma_{z})\rho_{a_{2}b_{2}c_{2}}^{++-}(k)]&=&\frac{1}{\sqrt{2}}
(\gamma_{1}^{(k)}+\gamma_{2}^{(k)}+\gamma_{3}^{(k)}-\gamma_{4}^{(k)})\nonumber\\
&=&\frac{1}{4\sqrt{2}}\mbox{Tr}\left(\rho_{ABC}^{k}+
(\rho_{ABC}^{T_{A}})^{k}+(\rho_{ABC}^{T_{B}})^{k}-(\rho_{ABC}^{T_{C}})^{k}\right),\nonumber\\
\mbox{Tr}[(\sigma_{z}\otimes\sigma_{z})\rho_{a_{2}b_{2}}^{++-}(k)]
&=&\beta_{1}^{(k)}=\frac{1}{2}\mbox{Tr}(\rho_{AB}^{T_{A}})^{k}\nonumber\\
\mbox{Tr}[(\sigma_{z}\otimes\sigma_{z})\rho_{a_{2}c_{2}}^{++-}(k)]
&=&\beta_{5}^{(k)}=\frac{1}{2}\mbox{Tr}\rho_{AC}^{k}\nonumber\\
\mbox{Tr}[(\sigma_{z}\otimes\sigma_{z})\rho_{b_{2}c_{2}}^{++-}(k)]
&=&\beta_{3}^{(k)}=\frac{1}{2}\mbox{Tr}\rho_{BC}^{k},
\end{eqnarray}
where $k=3,4,\cdots,d$.

Once Alice, Bob and Charlie complete all the four groups of
measurements, they can deduce the functions of the set of partial
transposition matrices. According to Eqs. \eqref{11}, \eqref{12},
\eqref{17}, \eqref{18}, \eqref{20} and \eqref{22}, they can get
\begin{eqnarray}\label{23}
&&\mbox{Tr}(\rho_{ABC}^{T_{A}})^{k}=2\mbox{Tr}[(\sigma_{z}\otimes\sigma_{z}\otimes\sigma_{z})\rho_{a_{1}b_{1}c_{1}}'(k)]
-2\sqrt{2}\mbox{Tr}[(\sigma_{z}\otimes\sigma_{z}\otimes\sigma_{z})\rho_{a_{2}b_{2}c_{2}}^{-++}(k)]\nonumber\\
&&\mbox{Tr}(\rho_{ABC}^{T_{B}})^{k}=2\mbox{Tr}[(\sigma_{z}\otimes\sigma_{z}\otimes\sigma_{z})\rho_{a_{1}b_{1}c_{1}}'(k)]
-2\sqrt{2}\mbox{Tr}[(\sigma_{z}\otimes\sigma_{z}\otimes\sigma_{z})\rho_{a_{2}b_{2}c_{2}}^{-+-}(k)]\nonumber
\end{eqnarray}
\begin{eqnarray}
&&\mbox{Tr}(\rho_{ABC}^{T_{C}})^{k}=2\mbox{Tr}[(\sigma_{z}\otimes\sigma_{z}\otimes\sigma_{z})\rho_{a_{1}b_{1}c_{1}}'(k)]
-2\sqrt{2}\mbox{Tr}[(\sigma_{z}\otimes\sigma_{z}\otimes\sigma_{z})\rho_{a_{2}b_{2}c_{2}}^{++-}(k)]\nonumber\\
&&\mbox{Tr}(\rho_{AB}^{T_{A}})^{k}=2\mbox{Tr}[(\sigma_{z}\otimes\sigma_{z})\rho_{a_{1}b_{1}}'(k)]
-2\mbox{Tr}[(\sigma_{z}\otimes\sigma_{z})\rho_{a_{2}b_{2}}^{-++}(k)]\nonumber\\
&&\mbox{Tr}(\rho_{AC}^{T_{A}})^{k}=2\mbox{Tr}[(\sigma_{z}\otimes\sigma_{z})\rho_{a_{1}c_{1}}'(k)]
-2\mbox{Tr}[(\sigma_{z}\otimes\sigma_{z})\rho_{a_{2}c_{2}}^{-++}(k)]\nonumber\\
&&\mbox{Tr}(\rho_{BC}^{T_{B}})^{k}=2\mbox{Tr}[(\sigma_{z}\otimes\sigma_{z})\rho_{a_{2}c_{2}}^{-++}(k)],
\end{eqnarray}
where $k=3,4,\cdots,d$. Combining the above equation with Eq.
\eqref{12}, they can determine the requisite eigenvalues and then
the set of negativites $\mathcal{N}_{(A-BC)}$,
$\mathcal{N}_{(B-AC)}$, $\mathcal{N}_{(C-AB)}$,
$\mathcal{N}_{(A-B)}$, $\mathcal{N}_{(A-C)}$ and
$\mathcal{N}_{(B-C)}$.

\section{discussions}
Our LOCC direct method is more parametrically efficient than the
LOCC quantum state tomography. For a $d$-dimensional tripartite
quantum state $\rho_{ABC}$, the quantum state tomography needs to
measure $d^{2}-1$ parameters. However, our direct method merely
requires to measure $4(d-2)+1$ parameters. Figure 2 shows the
number of parameters that need to be measured in quantum state
tomography (square) and our direct method (triangle) for the
$2\otimes 2\otimes 2$, $2\otimes 2\otimes 3$, $2\otimes 2\otimes
4$ and $2\otimes 3\otimes 3$ quantum states.
\begin{figure}[h]
\begin{center} \epsfig{figure=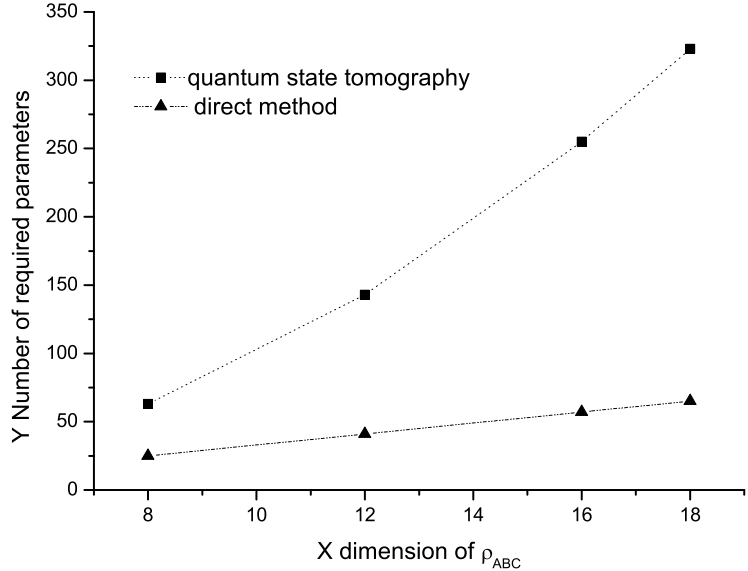,width=0.6\textwidth}
\caption{The number of required parameters in the LOCC quantum
state tomography (square) and our LOCC direct method (triangle)
for some lower dimensional tripartite quantum states.}
\end{center}
\end{figure}

In one-to-two party quantum communication, the observers possibly
care only about a part of the set of negativities. For example, in
the communication of Alice to Bob and Charlie, they only want to
know the negativities $\mathcal{N}_{(A-BC)}$, $\mathcal{N}_{(A-B)}$
and $\mathcal{N}_{(A-C)}$. In this case, the three observers need
only to make the first and the second group of measurements,
\emph{i.e.,} to measure $2d-3$ parameters. Then, by comparing the
tripartite relations and bipartite relations in Eqs. \eqref{10},
\eqref{11}, \eqref{17} and \eqref{18}, they can obtain the target
negativities. Similarly, if they measure the first and third group
of parameters (or the first and forth group of parameters), they can
obtain
$\{\mathcal{N}_{(B-AC)},\mathcal{N}_{(A-B)},\mathcal{N}_{(B-C)}\}$
(or
$\{\mathcal{N}_{(C-AB)},\mathcal{N}_{(A-C)},\mathcal{N}_{(B-C)}\}$)
in terms of corresponding relations.

As is known, the majorization criterion \cite{nak01} is stronger
than the entropic inequality. The criterion states that if
$\rho_{AB}$ is separable then
$\lambda(\rho_{AB})\prec\lambda(\rho_{A})$ and
$\lambda(\rho_{AB})\prec\lambda(\rho_{B})$, where $\lambda(\rho)$ is
the eigenvalue vector  of $\rho$.  The relation $x\prec y$ between
two n-dimensional vectors means that $\sum_{i=1}^{k}
x_{i}^{\downarrow}\leq\sum_{i=1}^{k} y_{i}^{\downarrow}, (1\leq
k\leq n-1)$ and $\sum_{i=1}^{n} x_{i}^{\downarrow}=\sum_{i=1}^{n}
y_{i}^{\downarrow}$, in which the symbol ``$\downarrow$" stands for
the decreasing order of the components of the vector. If the
dimensions of $x$ and $y$ are different, the smaller vector is
enlarged by appending extra zeros to equalize their dimensions. In
the tripartite system, Alice, Bob and Charlie may characterize the
entanglement in terms of a set of majorization criteria related to
the eigenvalue vectors $\lambda(\rho_{ABC})$, $\lambda(\rho_{AB})$,
$\lambda(\rho_{AC})$, $\lambda(\rho_{BC})$, $\lambda(\rho_{A})$,
$\lambda(\rho_{B})$ and $\lambda(\rho_{C})$. In order to testing the
set of criteria, they need to perform two groups of measurements, in
which  $2d-3$ parameters are measured. In the first group, they
implement the first part of the LOCC network shown in Fig.1 and then
measure the probabilities $P_{a_{1}b_{1}c_{1}}^{(k)}(ijl)$ for
$k=2,3,\cdots,d$. In the second group, they implement the whole LOCC
network in which all the controlled gates in the second part are
chosen to be controlled-$R_{+}$, and then measure the probabilities
$\rho_{a_{2}b_{2}c_{2}}^{+++}(k)$ for $k=3,4,\cdots,d$. After
completing these measurements, they can get two groups of relations,
by which they can obtain the target eigenvalue vectors.

However, in general, the majorization criterion is weaker than the
PPT criterion. Alice, Bob and Charlie can use a set of PPT criteria
to detect some bound entangled states which cannot be detected by
the corresponding majorization criteria. The D\"{u}r-Cirac-Tarrach
(DCT) state is just such a kind of quantum state, which takes the
form \cite{dct99}
\begin{eqnarray}\label{24}
 \rho_{DCT} &=&
 \sum_{\sigma=\pm}\lambda_{0}^{\sigma}\proj{\Psi_{0}^{\sigma}}+
 \sum_{k=01,10,11}\lambda_{k}(\proj{\Psi_{k}^{+}}+\proj{\Psi_{k}^{+}}),
\end{eqnarray}
where
$\ket{\Psi_{k}^{\pm}}=(\ket{k_{1}k_{2}0}\pm\ket{\overline{k}_{1}\overline{k}
_{2}1})/\sqrt{2}$, with $k_{1}$ and $k_{2}$ as the binary digits of
$k$ and $\overline{k}_{i}$ as the flipped $k_{i}$. When the
parameters are chosen to be $\lambda_{0}^{+}=\frac{1}{3}$,
$\lambda_{0}^{-}=\lambda_{10}=0$ and
$\lambda_{01}=\lambda_{11}=\frac{1}{6}$, the corresponding quantum
state is a bound entangled state and its matrix form reads
\begin{eqnarray}\label{25}
  \rho_{bound} =
 \left(%
\begin{array}{cccccccc}
  \frac{1}{6} & 0 & 0 & 0 & 0 & 0 & 0 & \frac{1}{6} \\
  0 & \frac{1}{6} & 0 & 0 & 0 & 0 & 0 & 0 \\
  0 & 0 & \frac{1}{6} & 0 & 0 & 0 & 0 & 0 \\
  0 & 0 & 0 & 0 & 0 & 0 & 0 & 0 \\
  0 & 0 & 0 & 0 & 0 & 0 & 0 & 0 \\
  0 & 0 & 0 & 0 & 0 & \frac{1}{6} & 0 & 0 \\
  0 & 0 & 0 & 0 & 0 & 0 & \frac{1}{6} & 0 \\
  \frac{1}{6} & 0 & 0 & 0 & 0 & 0 & 0 & \frac{1}{6} \\
\end{array}%
\right)
\end{eqnarray}
in the computational basis $\{\ket{ijl}\}$. The method for
generating and detecting the DCT bound entanglement was given by
Hyllus \emph{et al.} \cite{hab04}. Here, we redescribe the detection
procedure with our LOCC direct method. It is assumed that Alice, Bob
and Charlie share a number of copies of quantum state $\rho_{bound}$
which is unknown to the three observers. After performing the four
groups of measurements with the network shown in Fig.1, they can
obtain theoretically the data listed in Table 1.
\begin{table}[h]
\begin{center}
\begin{tabular}{c|c c c c c c c}
\hline\hline$\begin{array}{cc}
   & k \\
  <\sigma_{z}^{\otimes 3}>_{(k)} &  \\
\end{array}$ & 2 &  3 &  4 &  5 &  6 &  7 &  8 \\\hline
 $<\sigma_{z}^{\otimes 3}>_{(k)}^{a_{1}b_{1}c_{1}}$ &$$ $\frac{2}{9}$ $$ &
  $\frac{7}{144}$ &  $\frac{17}{1296}$ &  $\frac{19}{5284}$ &  $\frac{51}{46656}$ &
   $\frac{67}{186624}$ &  $\frac{197}{1679616}$ \\
 $<\sigma_{z}^{\otimes 3}>_{(k)}^{-++}$ & / &  $\frac{5}{144\sqrt{2}}$ &
  $\frac{13}{1296\sqrt{2}}$ &  $\frac{17}{5184\sqrt{2}}$ &  $\frac{49}{46656\sqrt{2}}$ &
   $\frac{65}{186624\sqrt{2}}$ &  $\frac{193}{1679616\sqrt{2}}$ \\
 $<\sigma_{z}^{\otimes 3}>_{(k)}^{-+-}$ & / &  $\frac{1}{48\sqrt{2}}$ &
  $\frac{7}{1296\sqrt{2}}$ &  $\frac{7}{5184\sqrt{2}}$ &  $\frac{19}{46656\sqrt{2}}$ &
   $\frac{23}{186624\sqrt{2}}$ &  $\frac{67}{1679616\sqrt{2}}$ \\
 $<\sigma_{z}^{\otimes 3}>_{(k)}^{++-}$ & / &  $\frac{1}{48\sqrt{2}}$ &
  $\frac{7}{1296\sqrt{2}}$ &  $\frac{7}{5184\sqrt{2}}$ &  $\frac{19}{46656\sqrt{2}}$ &
   $\frac{23}{186624\sqrt{2}}$ &  $\frac{67}{1679616\sqrt{2}}$ \\ \hline\hline
\end{tabular}
\caption{Theoretical values of the four groups of parameters for
the quantum state $\rho_{bound}$ in our LOCC direct method.}
\label{tab1}
\end{center}
\end{table}
Combining these data with Eq. \eqref{23}, the three observers can
deduce that $\rho_{bound}^{T_{A}}$ is negative and other partial
transposition matrices is semi-positive. Based on the negative
eigenvalue, they can obtain further the
$\mathcal{N}_{A-BC}(\rho_{bound})=\frac{1}{6}$. Here, it is noted
that only when an infinite ensemble of identically prepared output
state is given can Alice, Bob and Charlie determine the parameter
precisely. When a finite ensemble is given, they can only
determine the parameter approximately. Therefore, in order to
obtain these parameters with high fidelity, they need to run the
network many times. Especially, for $k$ is bigger, they need to
implement the network even more times.

  Although our method that makes use of bipartite entanglement measures
is limited to characterize partially the tripartite quantum state,
it can occasionally detect the genuine tripartite entanglement in
some specific cases. For example, in the case of three-qubit quantum
state, when $\mathcal{N}_{AB}=0$, $\mathcal{N}_{AC}=0$ and
$\mathcal{N}_{A-BC}>0$, we can deduce that the entanglement in the
bipartite splitting $A-BC$ is actually the genuine tripartite
entanglement among Alice, Bob and Charlie. This is because that the
negativities $\mathcal{N}_{AB}$ and $\mathcal{N}_{AC}$ can
characterize the two-qubit entanglement sufficiently. When the two
entanglements are zero, the residual entanglement
$\mathcal{N}_{A-BC}$ must be the tripartite entanglement. The
quantum state $\rho_{bound}$ in Eq. (25) is just the case.
Similarly, when $\mathcal{N}_{AB}=0, \mathcal{N}_{BC}=0$ and
$\mathcal{N}_{B-AC}>0$ (or $\mathcal{N}_{AC}=0, \mathcal{N}_{BC}=0$
and $\mathcal{N}_{C-AB}>0$), we are also able to judge that
$\mathcal{N}_{B-AC}$ (or $\mathcal{N}_{C-AB}$) is of tripartite
entanglement. Certainly, for a general three-qubit mixed state,
whether or not the tripartite entanglement exists cannot be
detected,  simply because a well-defined tripartite entanglement
measure is still unavailable, though a lot of efforts have been
made.

Two problems are worth to remark in our LOCC method. First, the
set of negativities  can only quantify some aspects of the
entanglement in a tripartite system. There exists some tripartite
entangled state \cite{abl01} that is PPT with respect to any of
the bipartite splitting. Moreover, as pointed by Hyluss \emph{et
al.} \cite{hab04}, there is a potential problem in practical
application of the direct methods, \emph{i.e.}, how to implement
effectively the controlled quantum gates, especially the
controlled-swap gate \cite{hfc95}. The solution of this problem
relies on the quantum technology that is currently being
developed.

\section{conclusions}
In this paper, we have generalized the direct approach of Hyllus
\emph{et al.} \cite{hab04} and proposed an LOCC method for detecting
a set of negativities in an arbitrary given tripartite quantum
state. The main task for the three observers is to measure
$4(d-2)+1$ parameters via three local networks supplemented by a
classical communication. Comparing with the LOCC quantum state
tomography which requires to measure $d^{2}-1$ parameters, our LOCC
method is more efficient. Moreover, our LOCC method does not require
the observers to perform the SPA of partial transposition maps,
which supports the Carteret's opinion \cite{car05} on the
three-party scenario, \emph{i.e.}, it is not the only way that they
make the quantum state undergo the partial transposition map, if
Alice, Bob and Charlie want to measure the function of the partial
transposition of $\rho_{ABC}$.

\section*{ACKNOWLEDGMENTS}
The work was supported by the RGC grant of Hong Kong under No.
HKU7045/05P, the URC fund of HKU, NSF-China grants under Nos.
10429401, 60325416 and 60328407, and the Special Foundation for
State Major Basic Research Program of China under grant No.
G2001CB309500.

\end{document}